\documentclass[a4paper]{article}
\usepackage{atmohead2013}
\usepackage[english]{babel}
\usepackage{amsmath}
\usepackage{amssymb}
\usepackage{amsthm}
\usepackage[utf8]{inputenc}

\title{Early attempts at atmospheric simulations for the Cherenkov Telescope Array}

\shorttitle{Early attempts at atmospheric simulations for the Cherenkov Telescope Array}

\authors{
Rulten, C.B.$^{1,2}$ and
Nolan, S.J.$^{1}$ for the CTA consortium.
}

\afiliations{
$^1$ Physics Department, Durham University, South Road, Durham, County Durham, DH1 3LE \\
$^{2}$ now at: LUTH, L'Observatorie de Paris, 5, place Jules Janssen, 92195 Meudon, FRANCE \\
}

\email{cameron.rulten@obspm.fr}

\abstract{The Cherenkov Telescope Array (CTA) will be the world's first observatory for detecting gamma-rays from astrophysical phenomena and is now in its prototyping phase with construction expected to begin in 2015/16. In this work we present the results from early attempts at detailed simulation studies performed to assess the need for atmospheric monitoring. This will include discussion of some lidar analysis methods with a view to determining a range resolved atmospheric transmission profile. We find that under increased aerosol density levels, simulated gamma-ray astronomy data is systematically shifted leading to softer spectra. With lidar data we show that it is possible to fit atmospheric transmission models needed for generating lookup tables, which are used to infer the energy of a gamma-ray event, thus making it possible to correct affected data that would otherwise be considered unusable.}
\newcommand*{\myparskip}{\vspace{\smallskipamount}}

\begin{document}
\pagenumbering{arabic}\maketitle

\section{Introduction}
\noindent The Cherenkov Telescope Array (CTA) aims to increase its flux sensitivity by an order of magnitude compared to existing ground-based gamma-ray telescopes \cite{bib:mcpaper}. This will be achieved using Cherenkov telescopes of 3 different sizes, a large size telescope (LST) $\mathrm{\sim 23\thickspace m}$ diameter, a medium size telescope (MST) $\mathrm{\sim 12\thickspace m}$ diameter and a small size telescope (SST) $\mathrm{\sim 4\thickspace m}$ diameter. In order to achieve such sensitivity gains it is important to minimise the systematic uncertainty in derived flux and energy resolution. Imaging atmospheric Cherenkov telescopes are calorimetric by nature and as such a good knowledge of atmospheric conditions is required at the telescope site. Atmospheric quality affects both shower development and the Cherenkov yield in two ways. Firstly, in the production of Cherenkov light atmospheric quality affects the vertical profile of the refractive index of the air and hence shower development. This variation is seasonal, and effects mid-latitudes worse than the tropics. However, the profile can be measured using radiosondes for example and any seasonal variation that might exist can be accounted for. It is also possible for high-level aerosols (e.g. clouds) to occur around shower maximum and so affect Cherenkov yield and image shape.

\myparskip

\noindent Secondly, poor atmospheric quality can also result in the loss of Cherenkov light. For example atmospheric quality affects Cherenkov light propagation through Rayleigh \& Mie scattering of the Cherenkov light, which can lower the brightness of an image in the camera for a shower of given energy and core distance. However, by using lidar measurements it is possible to derive a range-resolved probability of transmission (at the lidar wavelength) and adjust atmospheric models, needed in simulations used to reconstruct gamma-ray spectra, accordingly \cite{bib:nolan}.

\myparskip
\noindent This work highlights an early simulation study conducted using a hypothetical 97 telescope array to illustrate the effects of atmospheric quality on a reconstructed gamma-ray spectrum.

\myparskip
\noindent Finally, another motivational factor for a large observatory like CTA is the desire to maximise the duty cycle of the instrument. Thus being able to resurrect otherwise unusable data due to relatively poor atmospheric conditions becomes important.

\section{Technique}
\noindent For this work measurements of the atmosphere were recorded using an Easy-Lidar ALS450XT developed with and manufactured by Leosphere France. The Easy-Lidar ALS450XT was a monostatic bi-axial lidar that is now defunct. Table \ref{table1} highlights the specifications of the lidar used for this research.

\myparskip
\begin{table}[!h]
\begin{center}
{\footnotesize
\begin{tabular}{| l | r |}
\hline
Wavelength & 355 nm \\
Frequency & 10 Hz \\
Pulse Width & 5 ns \\
Energy/Pulse & 20 mJ \\
Range & 15 km \\
Resolution & 1.5 m \\
\hline
\end{tabular}
}
\end{center}
\caption{Specifications of the Leosphere Easy-Lidar ALS450XT.}   
\label{table1}
\end{table}

\begin{figure*}[!htb]
\begin{minipage}[l]{1.0\columnwidth}
\centering
\includegraphics[width=\textwidth]{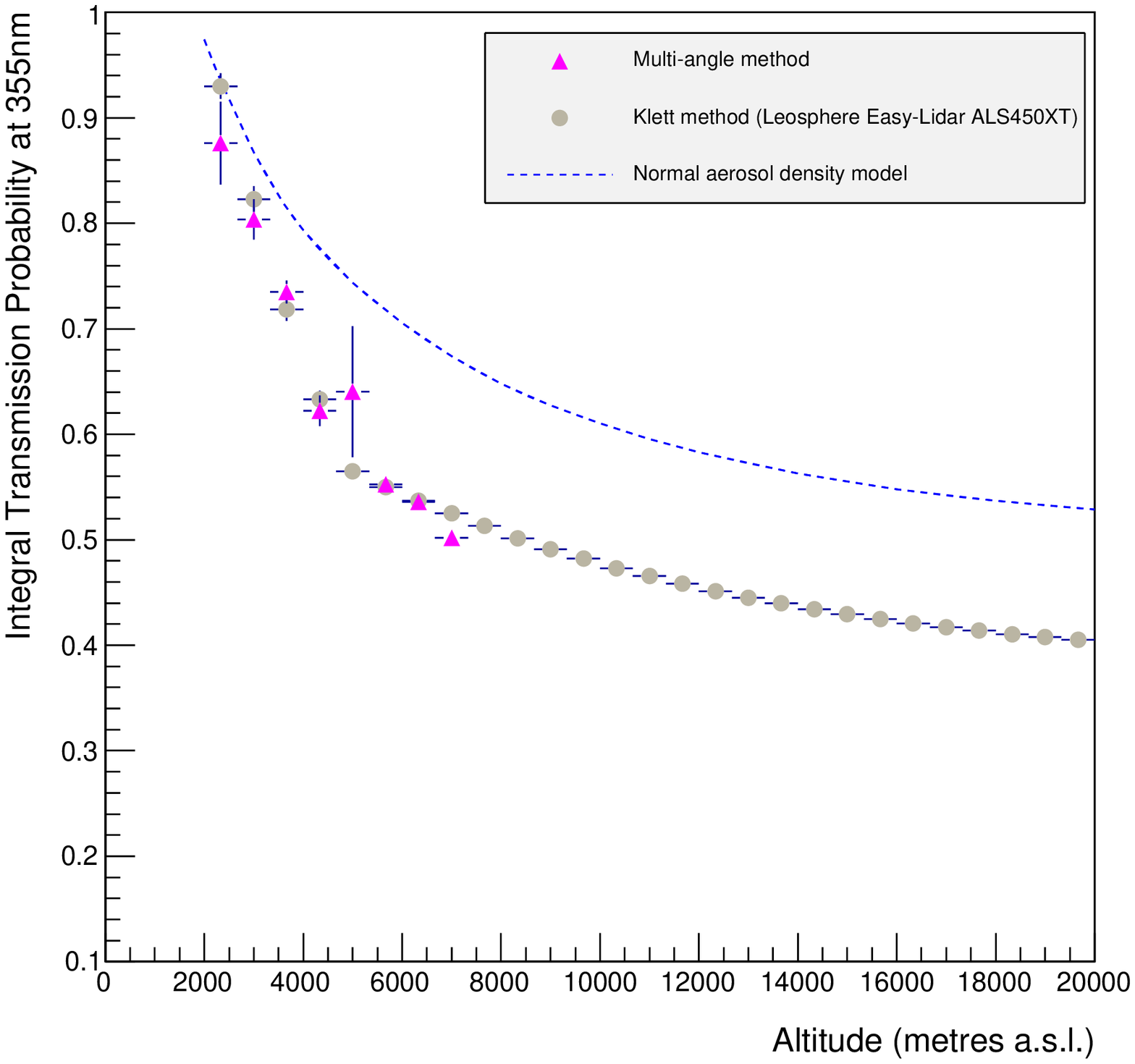}
\caption{The derived range-resolved atmospheric transmission up to $\mathrm{20\thickspace km}$ above sea level (a.s.l.) recorded at the H.E.S.S. site ($\mathrm{1800\thickspace m\thickspace a.s.l.}$) in Namibia on 15th August 2008. The pink triangles show the transmission derived using the multiangle method, the grey circles show the transmission derived using the Klett method and the blue dashed line shows the normal aerosol density transmission model widely used to characterise the measurement site.}
\label{figure:1}
\end{minipage}
\hfill{}
\begin{minipage}[r]{1.0\columnwidth}
\centering
\includegraphics[width=\textwidth]{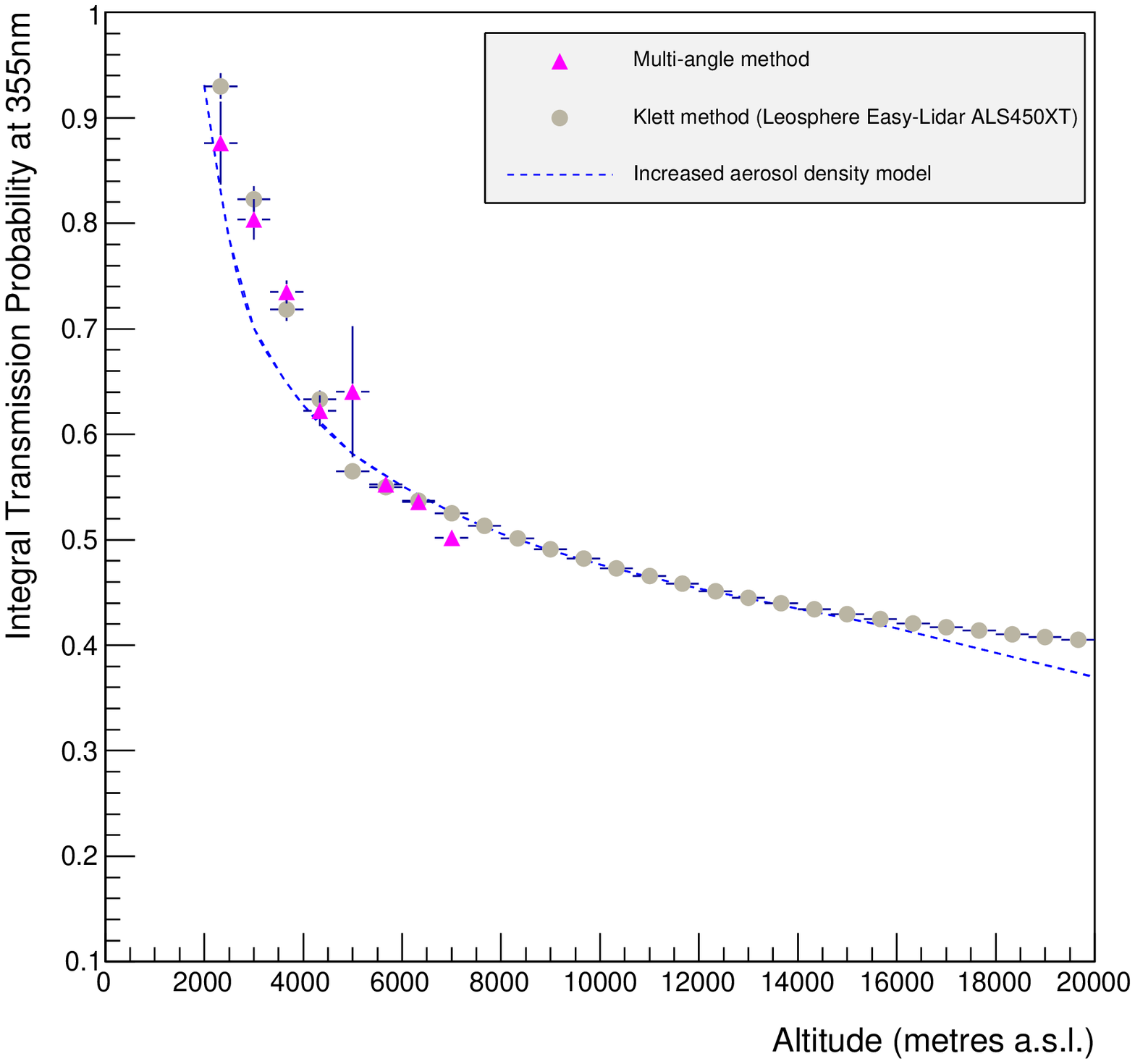}
\caption{The derived range-resolved atmospheric transmission up to $\mathrm{20\thickspace km}$ above sea level (a.s.l.) recorded at the H.E.S.S. site ($\mathrm{1800\thickspace m\thickspace a.s.l.}$) in Namibia on 15th August 2008. The pink triangles show the transmission derived using the multiangle method, the grey circles show the transmission derived using the Klett method and the blue dashed line shows the newly derived best-fit aerosol model found using MODTRAN.}
\label{figure:2}
\end{minipage}
\end{figure*}

\noindent The approach adopted for acquiring data involved pointing the lidar toward zenith and firing the laser to acquire a single atmospheric profile averaged over 600 laser shots. The transmission profile was derived from the lidar data using the Klett inversion method \cite{bib:klett} and the multiangle method \cite{bib:multi}. Using lidar data recorded on 15th August 2008 at the H.E.S.S. site ($\mathrm{23^{\circ}16.28'S 16^{\circ}30'E}$), when the atmospheric quality was perceived to be visibly poor, the range-resolved atmospheric transmission was derived as shown in Figure \ref{figure:1}.

\myparskip

\begin{figure*}[ht]
\begin{minipage}[l]{1.0\columnwidth}
\centering
\includegraphics[width=\textwidth ,height=0.35\textheight,clip]{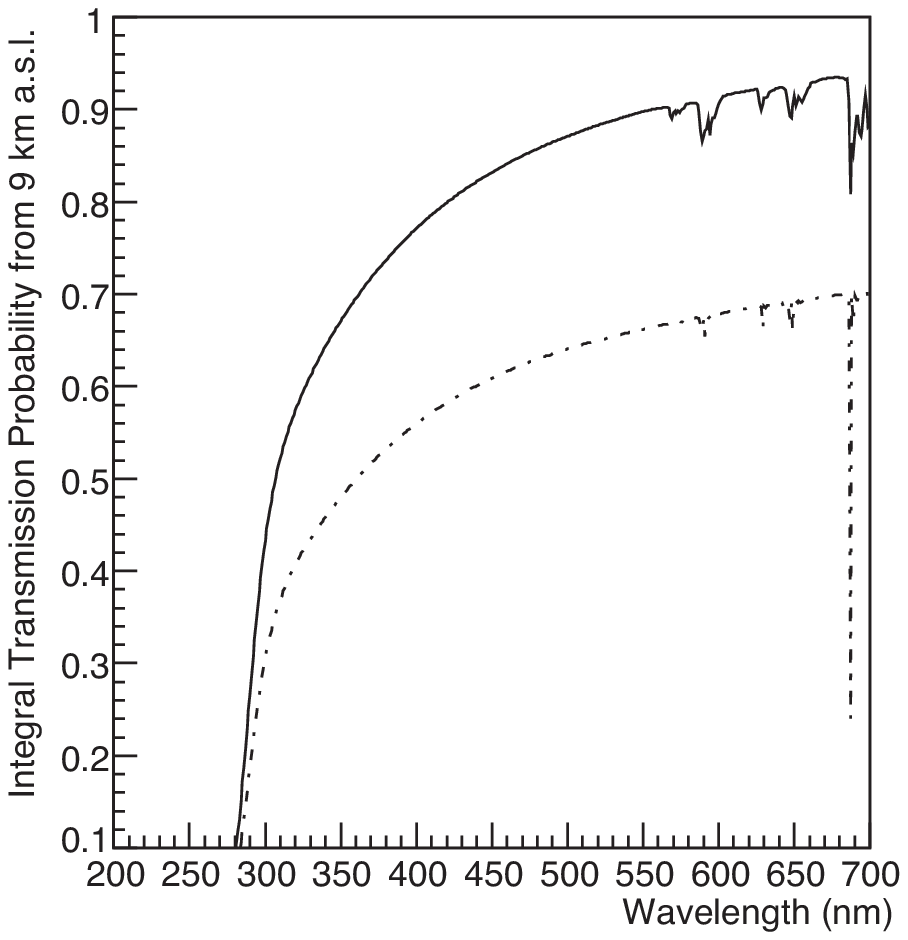}
\caption{Shown here is the probability of transmission from $\mathrm{9\thickspace km}$ versus wavelength for the normal aerosol density model (solid black line) widely used to characterise the measurement site, and the newly derived 'best-fit' aerosol model generated using MODTRAN (dash-dot black line) referred to in this work as the increased aerosol density model.}
\label{figure:3}
\end{minipage}
\hfill{}
\begin{minipage}[r]{1.0\columnwidth}
\centering
\includegraphics[width=\textwidth,height=0.32\textheight,clip]{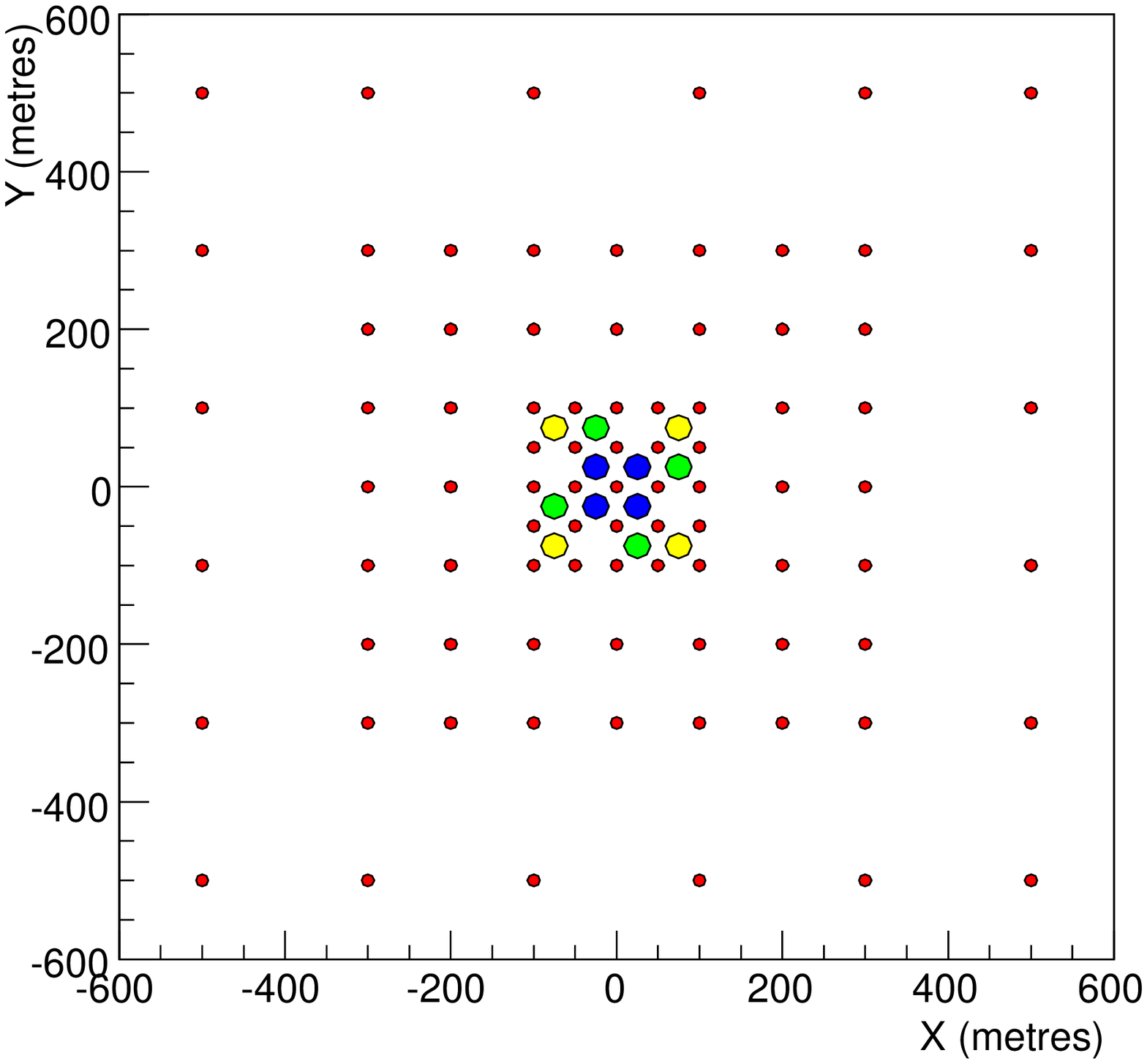}
\caption{Shown here is a scaled image of the simulated telescope layout as positioned on the ground. The large yellow, green and blue dots represent the large-sized $\mathrm{\sim 600\thickspace m^{2}}$ area telescopes and the smaller red dots represent the medium-sized telescopes.}
\label{figure:4}
\end{minipage}
\end{figure*}

\noindent The standard 'desert-dust' transmission model \cite{bib:modtran} is widely used to characterise the measurement site. For convenience this is referred to as the normal aerosol density model in this work. The normal aerosol density transmission model (dashed blue line) shown in Figure \ref{figure:1} is for a spectrum of wavelengths (100 nm to - 1 mm) and is derived using MODTRAN (version 4) \cite{bib:modtran}. All models of vertical aerosol density (plus the molecular absorption and scattering) cited within this paper have been simulated using MODTRAN. As already noted, the lidar derived transmission profile is for a single wavelength only. However, by adjusting the aerosol density within MODTRAN (see Figure \ref{figure:3}) to match the transmission recorded at the lidar wavelength, it is possible to find a best fit aerosol profile (dashed blue line in Figure \ref{figure:2}). The use of a multiwavelength lidar would make it possible to achieve a much better fit with the MODTRAN models.

\myparskip

\noindent Atmospheric air shower simulations were conducted using CORSIKA and the telescope simulations were conducted using the \emph{sim\_telarray} software package \cite{bib:simtelarray}. 20 million gamma-ray showers between 5 GeV and 2 TeV, with a differential power-law spectrum with slope $\mathrm{E^{-2}}$ at zenith were produced. A hypothetical CTA telescope array comprising 97 telescopes was used and the atmospheric transmission models were folded in at this telescope simulation stage. In total, 2 different telescope response databases were produced; one for good atmospheric conditions i.e. produced with the normal aerosol density transmission model widely used to characterise the measurement site and one for poor atmospheric conditions i.e. produced using the increased aerosol density transmission model derived using lidar data recorded at the measurement site. Figure \ref{figure:4} illustrates the 97 telescope system comprising two different telescope types. This includes 12 large-sized parabolic dish telescopes each with a mirror area of $\mathrm{\sim 600\thickspace m^{2}}$, 4093 pixels at their primary focus and a 5 degree field of view. In addition, there are 85 medium-sized Davies-Cotton dish telescopes each with a mirror area of $\mathrm{\sim 100\thickspace m^{2}}$, 1735 pixels at their primary focus and a 7 degree field of view. It should be noted that such a telescope array is unlikely to be built for the CTA as it would be too costly and does not necessarily provide optimal sensitivity performance of the broadband energy range from 10 GeV to above 100 TeV. Furthermore, the photomultiplier quantum efficiency for both of the simulated telescopes was increased by 50\% compared to the Photonis XP2960, and a 3 pixel trigger threshold was set requiring a minimum signal of 5.3 photoelectrons (within a given sector of the camera). Events which failed to meet this criteria were discarded from the analysis.

\myparskip
\begin{figure*}[ht]
  \begin{center}
    \includegraphics[width=0.9\textwidth, height=0.37\textheight,clip]{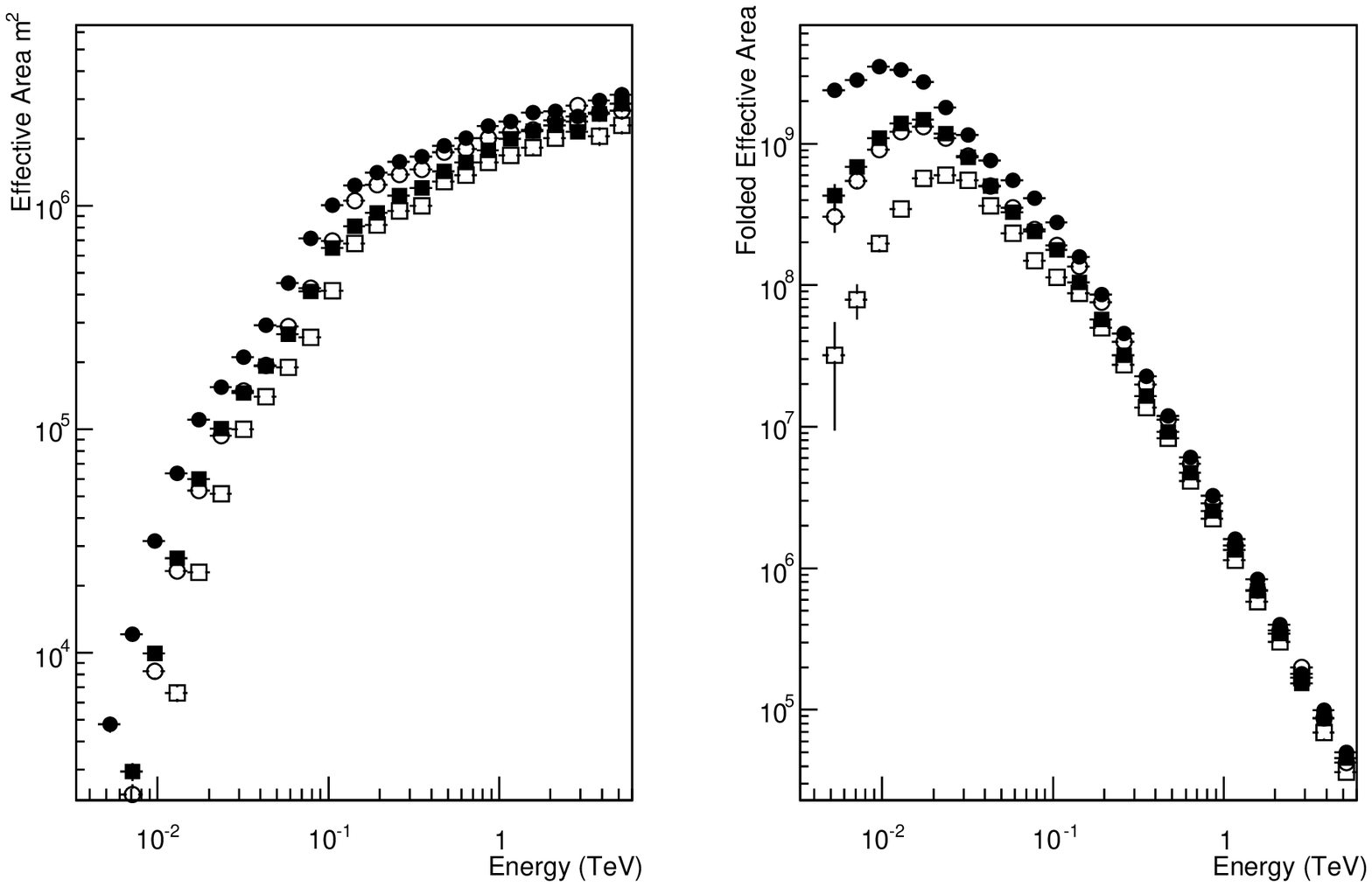}
  \caption{\label{figure:5} {\it Left:} The effective area for triggering is shown for a database of gamma$-$ray showers at zenith folded with a telescope simulation that is computed with different atmospheric models to reflect the real atmospheric quality measured with a lidar. Shown here is the resulting effective area for the normal aerosol density model (filled circles) and the increased aerosol density model (filled squares). In addition a cut of at least 2 triggering telescopes, with a minimum of 4 signal tubes is applied to the data and the resulting effective area after this cut is shown for the normal aerosol density model (open circles) and the increased aerosol density model (open squares).  {\it Right:} Shown here is the effective area (seen in the Left panel) folded with a power-law spectrum (Crab type slope of $\mathrm{E^{-2.45}}$) in order to illustrate the energy threshold of the system, which is located at the peak of each distribution.}
\end{center}
 \end{figure*}

\noindent Figure \ref{figure:5} (Left panel) shows the effective area for the simulated telescope system derived from both databases of telescope simulations (normal and increased aerosol density levels) for both events which trigger, and events which pass loose quality cuts of at least 2 triggering telescopes, with a minimum of 4 signal tubes in each camera. In addition, the right panel of Figure \ref{figure:5} (Right panel) shows the effective areas folded with a power-law spectrum whose slope is $\mathrm{E^{-2.45}}$ (a Crab-like spectrum), to indicate the threshold energy of the system \footnote{There are other conventions for determining the energy threshold.}. For the normal aerosol density dataset, the triggering threshold energy is 10 GeV, rising to 20 GeV post the loose cuts.

\myparskip
\noindent Whereas for the increased aerosol density dataset the minimum triggering threshold is 20 GeV rising to 30 GeV post the loose cut. Initially it appears that such a change in atmospheric quality has little effect on the simulation, but this is misleading and doesn't illustrate the picture completely. In order to perform spectroscopy, one must reconstruct the energy of an event. Typically a lookup table, derived from simulation, is used to infer the energy of the event. This energy is a function of the impact parameter (r) and the logarithm of the image brightness (S) and can be represented as a function E(r, S) \cite{bib:crabpaper}. However, S is a function of both energy and atmospheric transmission (T), represented as S(E, T). Thus the observed Cherenkov image brightness (size) is also dependent on atmospheric quality as well as energy. For the increased aerosol density dataset the size of an event, for a shower of given impact distance and energy, will be less compared to that using the normal aerosol density dataset as a larger number of Cherenkov photons are scattered away from the telescope field of view.

\section{Results} 
\noindent In order to test the effect of atmospheric quality on the simulated dataset,  a set of lookup tables for the reconstructed energy $\mathrm{E_{R}(r,S)}$ and the reconstructed effective area $\mathrm{A_{R}(E)}$ were generated from the simulation database. This was done for both a normal aerosol density level and an increased aerosol density level. A test spectrum of 100,000 events each with a simulated energy E (following a power-law spectral slope of $\mathrm{E^{-2.3}}$) was randomly drawn from each of the databases. Using the lookup tables, the $\mathrm{E_{R}(r,S)}$ and $\mathrm{A_{R}(E)}$ were derived for these events and a reconstructed differential spectrum was generated for three specific combinations of simulation data and lookup tables highlighted in Table \ref{table:2}.

\myparskip
\begin{table*}[!h]
  \centering
  \begin{tabular}{|c|c|c|}
  \hline
  Case   &  Simulated data derived from & Lookup table derived from \\
   \hline 
    1 & Normal aerosol density database & Normal aerosol density database \\
    2 & Increased aerosol density database & Normal aerosol density database \\
    3 & Increased aerosol density database & Increased aerosol density database \\
    \hline
  \end{tabular}
  \caption{Reconstructed differential spectra were generated using the following three specific combinations of simulation data and lookup tables.}
  \label{table:2}
  \end{table*}

\begin{figure*}[th]
  \centering
  \includegraphics[width=0.9\textwidth]{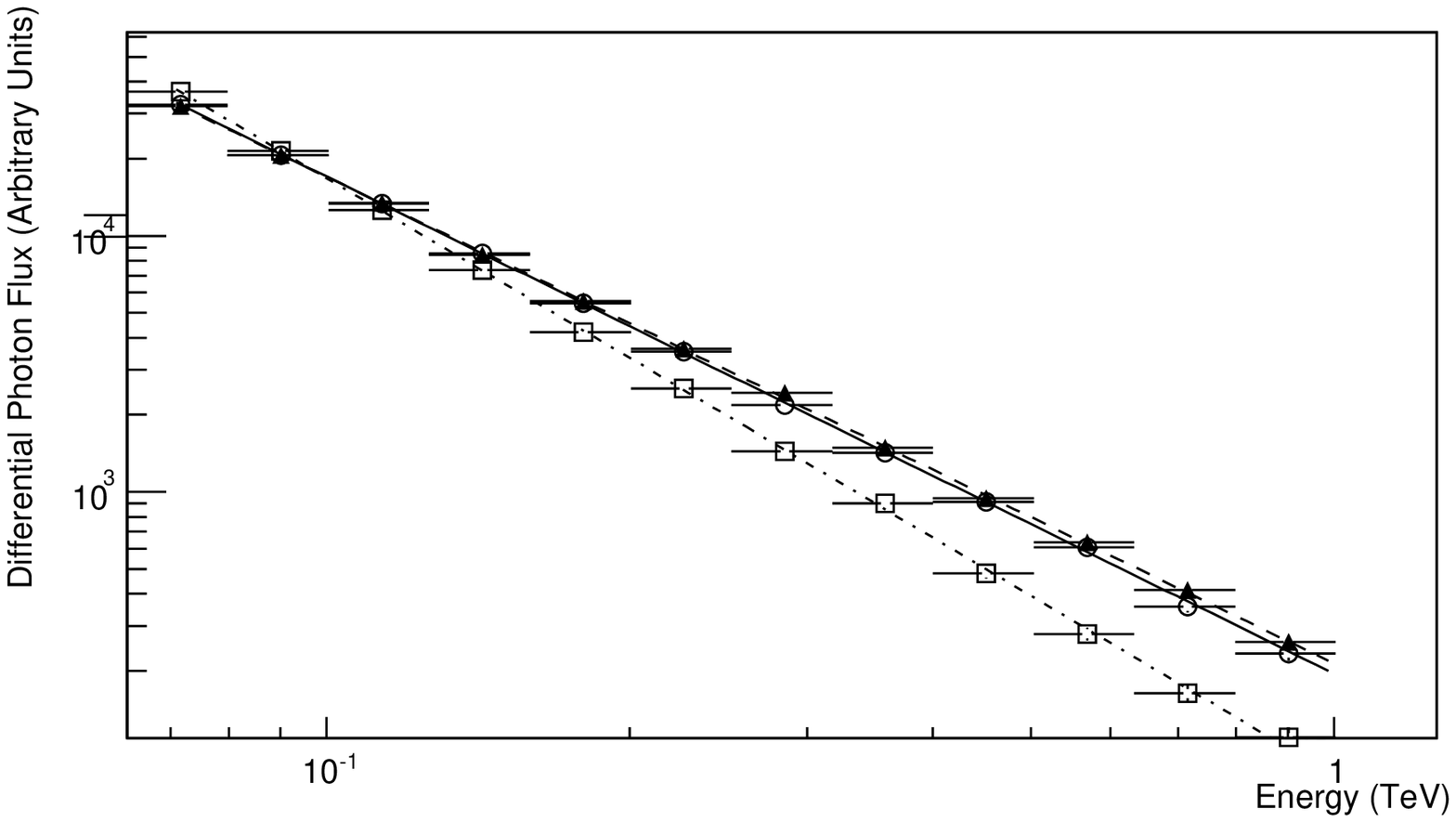}
  \caption{Shown here are the reconstructed differential spectra generated from lookup tables that were created from a random sample of 100,000 events following a spectral slope of $\mathrm{E^{-2.3}}$. This has been done for each of the different combinations of simulation data and lookup tables. The  open circles show the reconstructed differential spectrum for case 1, the open squares for case 2 and the filled triangles for case 3. By incorporating lidar data into the reconstruction, this allows a corrected spectrum to be formed (case 3) whose slope is very similar to that seen when the atmospheric quality is considered to have a normal aerosol density level (case 1).}
  \label{figure:6}
 \end{figure*}

\noindent Figure \ref{figure:6} shows the reconstructed spectra generated using the different combinations of simulation data and lookup tables.

\section{Discussion}
\noindent Figure \ref{figure:5} shows that without correction, changing atmospheric quality has a significant effect on the simulated telescope response which results in the reconstructed spectrum being systematically shifted. Dimmed shower images result in two factors causing this shift to occur; firstly the telescope triggering efficiency decreases around threshold as dimmed shower images fail to meet the triggering criteria, and secondly the energy for a given event is systematically reconstructed to a lower energy value. Table \ref{table:3} highlights the resulting fit to the data for a power law of form $\mathrm{\frac{dN}{dE}=I_{o}E^{-\alpha}}$ where $\mathrm{dN/dE}$ is the differential photon flux in  events $\mathrm{TeV^{-1}}$.
\begin{table}[!h]
  \centering
  \begin{tabular}{|c|c|c|}
  \hline
  Case   &  $\mathrm{I_{o}}$ & $\alpha$ \\
   &events $\mathrm{TeV^{-1}}$&\\
   \hline 
1 &198$\pm$3&  1.93$\pm$0.01\\
2 &77$\pm$2&  2.34$\pm$0.01\\
3 &210$\pm$3&  1.91$\pm$0.01\\
\hline
  \end{tabular}
  \caption{Results of fitting a power-law $\mathrm{\frac{dN}{dE}=I_{o}E^{-\alpha}}$ to the reconstructed spectra for each of the cases highlighted in Table \ref{table:2}. The quoted errors are statistical only.}
  \label{table:3}
  \end{table}

\noindent The power-law fit values highlighted in Table \ref{table:3} show that correcting for a changed atmospheric quality results in a power-law index similar to what is expected when atmospheric quality at the site is considered to be normal, or at least the aerosol density level is considered to be normal. Thus using a lidar to measure changing aerosol density levels and hence atmospheric quality is a useful tool for correcting data in ground-based gamma-ray astronomy. However, it should be noted that transmission values calculated using a single-scattering lidar like the one used in this work, are reported to have systematic errors of approximately $\mathrm{30 \%}$ \cite{bib:auger}. Thus there is a desire to test whether Raman lidars which have a much lower associated systematic error ($\mathrm{\sim 5\%}$) on the derived range-resolved transmission can be used within CTA for active atmospheric calibration.

\section{Conclusion}
\noindent Currently within the field of ground-based gamma-ray astronomy, atmospheric quality is accounted for by monitoring the background cosmic-ray trigger rates and data with sub-standard atmospheric quality is discarded \cite{bib:nolan}. This work shows that it is possible to use a lidar to take in-situ atmospheric measurements in order to derive the probability of transmission at a wavelength close to the maximum for Cherenkov light production. A model of atmospheric transmission for a spectrum of wavelengths is then fitted to the lidar data and used within simulations to produce lookup tables that better reflect the actual atmospheric quality. Correcting for changing atmospheric quality in such a way can increase the lifetime of an observatory like CTA. In addition, such an active atmospheric calibration method helps to lower any systematic uncertainty on the derived flux.

\myparskip
\par{\noindent {\it Acknowledgements}  The authors would like to acknowledge the support of their host institutions and also the support from the Science and Technology Facilities Council of the UK. In addition, the authors gratefully acknowledge support from the agencies and organizations listed on this page: http://www.cta-observatory.org/?q=node/22}

\end{document}